\documentclass[noshowpacs,amsmath,twocolumn,superscriptaddress,8pt,aps,prb]{revtex4-1} 
\usepackage{setspace}
\usepackage{amsmath}
\usepackage{breqn}
\usepackage{graphicx}
\usepackage[nearskip,margin = 0pt]{subfig}
\usepackage{verbatim} 
\usepackage{amsfonts}
\usepackage{amssymb}
\usepackage{makecell}
\usepackage{epstopdf}
\usepackage{xcolor}
\usepackage[resetlabels,labeled]{multibib}
\DeclareGraphicsExtensions{.pdf,.eps,.png,.jpg,.mps}

\def\vx{\ensuremath\textbf{x}}
\def\vE{\ensuremath\textbf{E}}
\def \vPhi{\ensuremath{\Phi}}
\def \vk{\ensuremath\textbf{k}}

\begin{document}
\title{Low-overhead distribution strategy for simulation and optimization of large-area metasurfaces}
\author{Jinhie Skarda$^{1,*}$, Rahul Trivedi$^{1,2,3,*,\dagger}$, Logan Su$^{1,*}$, Diego Ahmad-Stein$^{1}$, Hyounghan Kwon$^{1}$, \\Seunghoon Han$^{4}$, Shanhui Fan$^{1}$, and Jelena Vu\v{c}kovi\'{c}$^{1,\dagger}$\\
\vspace{+0.05 in}
$^1$E. L. Ginzton Laboratory, Stanford University, Stanford, CA 94305, USA.\\
$^2$Max-Planck-Institut für Quantenoptik, Hans-Kopfermann-Str.~1, 85748 Garching, Germany.\\
$^3$Department of electrical and computer engineering, University of Washington, Seattle, 98195. \\
$^4$Samsung Advanced Institute of Technology, Samsung Electronics, Suwon-si, Gyeonggi-do 443-803, Republic of Korea\\
*These authors contributed equally to this work.\\
$^{\dagger}$Corresponding authors: rtriv@uw.edu, jela@stanford.edu}

 \maketitle

{\bf 
Fast and accurate electromagnetic simulation of large-area metasurfaces remains a major obstacle in automating their design. In this paper, we propose a metasurface simulation distribution strategy which achieves a linear reduction in the simulation time with the number of compute nodes. Combining this distribution strategy with a GPU-based implementation of the Transition-matrix method, we perform accurate simulations and adjoint sensitivity analysis of large-area metasurfaces. We demonstrate ability to perform a distributed simulation of large-area metasurfaces (over $600 \lambda \times 600 \lambda$), while accurately accounting for scatterer-scatterer interactions significantly beyond the locally periodic approximation.}

Being able to achieve full phase control of optical fields is a central challenge in optical engineering, with diverse applications in imaging, sensing, augmented, and virtual reality systems \cite{lee2019prospects,berkovic2012optical}. The past decades have seen a rapid development of metasurface-based optical elements that exploit collective scattering properties of subwavelength structures for phase-shaping the incoming fields and are significantly more compact and integrable when compared to the conventional refractive optical elements \cite{chen2020flat,zhou2020flat,colburn2018metasurface,horie2016wide,kwon2020single,lee2018metasurface,li2021inverse}. The most commonly adopted metasurface-design strategy proceeds in two steps --- \emph{first}, a library of periodic meta-atoms with varying transmission amplitudes and phases is generated by varying a few geometric parameters specifying the meta-atom. \emph{Next}, an aperiodic meta-surface is generated by laying out the periodic meta-atoms corresponding to the target spatially-varying phase profile \cite{mcclung2020will,arbabi2020increasing,pestourie2020active, zhan2016low, aieta2012aberration, arbabi2016multiwavelength,devlin2016broadband,fan2018silicon}. This approach suffers from two major limitations --- \emph{first}, the resulting metasurface should be almost periodic, and thus this strategy cannot be used for reliably designing rapidly varying phase-profiles. \emph{Second}, generating the metasurface library becomes increasingly difficult for multi-functional design problems. For instance, while it is usually not difficult to generate a library for designing a simple phase-mask operating at a few operating modes \cite{shi2018single,arbabi2016multiwavelength,khorasaninejad2016polarization}, it becomes increasingly difficult to scale up the number of modes since the same metasurface is required to simultaneously satisfy multiple design conditions corresponding to the different input modes.

Fully automating design of metasurfaces can provide a potential solution to this problem. Gradient based optimization has been successful in designing integrated optical elements that are more compact, robust and high performing than their classical counterparts \cite{Molesky:2018:NP,yang2020inverse,wang2011robust,hughes2018adjoint,sapra2020chip,dory2019inverse,piggott2015inverse,su2020nanophotonic}. A key ingredient in these approaches is the ability to rapidly simulate the full electromagnetic structure. This presents a challenge for metasurface designs, since practical metasurfaces could be approximately $10^2$ - $10^3$ $\lambda$ in the linear dimension, making it impractical to use general-purpose electromagnetic solvers such as Finite-Difference Time-Domain (FDTD) \cite{taflove2000computational}, Finite-Difference Frequency-Domain (FDFD) \cite{rumpf2012simple}, or Finite Element Method (FEM) \cite{reddy2019introduction}. Inverse-design approaches that use discrete general-purpose electromagnetic solvers to simulate and design the full surface are limited to small design areas or a small number of optimization iterations\cite{camayd2020multifunctional, mansouree2020multifunctional}, or restrict the parameter space through a specific symmetry that allows for fast simulations \cite{christiansen2020fullwave,lin2021computational, chung2020high}. Consequently, nearly all the current methods for inverse-designing large-scale 3D metasurfaces rely on approximate electromagnetic simulations of the metasurface locally using either periodic or radiation boundary conditions \cite{li2021inverse,lin2019topology,pestourie2018inverse,chung2020tunable,sell2017periodic,phan2019high,lin2019overlapping,sell2017large,bayati2021inverse,zhelyeznyakov2021deep,jiang2019global,jiang2019free,byrnes2016designing}, which do not accurately account for interactions between different meta-atoms. These approaches are thus fundamentally limited to designing metasurfaces with slow phase variations due to the implicit local approximation. A coupled-mode formalism can also be applied for metasurface simulation and optimization \cite{doi:10.1021/acsphotonics.1c00100} but this approach is not guaranteed to yield exact fields, particularly for metasurfaces with multiple low quality-factor modes.

In this paper, we propose and demonstrate a numerically accurate simulation strategy that can be used to design and analyze large-area metasurfaces. Our strategy relies on a distribution of the simulation method where the simulation time scales linearly with the compute resources. This is achieved by a Nyquist-sampling decomposition of the fields incident on the metasurface, similar to that used recently to characterize the discrete impulse response of aperiodic metasurfaces \cite{torfeh2020modeling}. Our distribution strategy, by ensuring minimal communication between compute nodes, allows for a linear reduction in the simulation time with the number of compute nodes, indicating that arbitrarily large metasurfaces can be simulated in reasonable time with sufficiently large number of compute nodes. On each compute node, we implement a GPU-based transition-matrix (T-matrix) simulation\cite{zhan2019controlling, zhan2018inverse, zhelyeznyakov2020design}. Though there are GPU-optimized FDTD implementations that allow fast simulation of unit-cells up to 100 $\lambda$ $\times$ 100 $\lambda$ \cite{hughestiny3d}, these approaches do not currently provide a low-overhead means of parallel simulation distribution. We demonstrate numerically accurate simulations of metasurfaces of size 1mm $\times$ 1 mm at a wavelength of $1.55 \mu$m (about $645 \lambda \times 645 \lambda$) on a cluster of 48 GPU nodes. Finally, we demonstrate the ability to efficiently compute the gradients with respect to both the geometry and the positions of the meta-atoms, thus enabling the application of optimization-based design to large-scale metasurfaces.

\noindent\textbf{Low-Overhead Multi-GPU Simulation Strategy.}

To simulate millimeter-scale metasurfaces, it is essential to parallelize the simulation method across multiple compute nodes. In order to be scalable, however, this parallelization scheme should introduce only a modest communication overhead in the simulation as this communication overhead can potentially offset any time savings achieved due to the parallelization (\cite{tang2020parallelized, hermann2010multi, dziekonski2017communication}).

\begin{figure*}[h!]
\centering
\includegraphics[width=\linewidth]{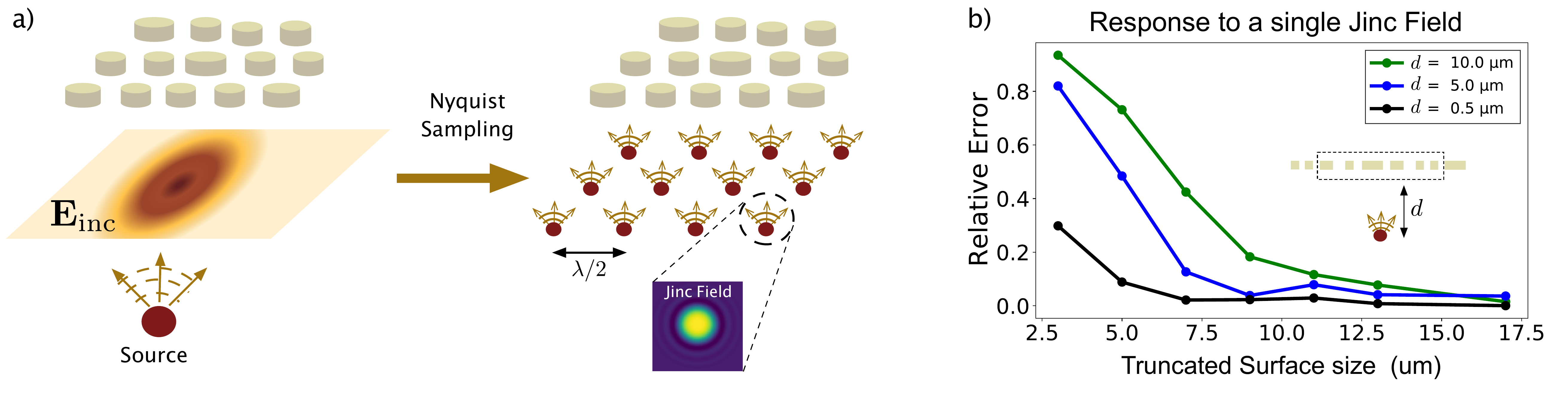}
\captionsetup{singlelinecheck=no}
\caption{{\bf{Nyquist sampling of bandlimited incident field.}} (\textbf{a}) Schematic of Nyquist sampling of the incident electric field, which is bandlimited because it is propagating. (\textbf{b}) Percent error in scattered field power versus spatial-extent of metasurface included in the simulation for a single jinc source placed 10 $\mu$m (green), 5 $\mu$m (blue), and 0.5 $\mu$m (black) from the metasurface. The full metasurface is a 25 $\mu$m $\times$ 25 $\mu$m metasurface with focal length of 10 $\mu$m, and the surface size on the x-axis of this convergence plot refers to the spatial-extent around the center of this metasurface that is included in the simulation. The y-axis relative error is computed assuming the simulation including the full metasurface is the converged result.}
\label{fig:Fig2}
\end{figure*}

For metasurface simulations, however, by utilizing the property that the incident fields generated by far-field sources will be within the light-cone in the $\textbf{k}-$space, a parallelization strategy can be devised that requires minimal communication between the compute nodes. The fundamental principle behind this parallelization is to represent the bandlimited incident field by its samples using the Nyquist sampling theorem \cite{landau1967sampling}. More precisely, consider an incident field propagating along the $z-$direction --- the transverse polarization of this field, $\textbf{E}_{inc}^T(x, y, z)$ at any $z$ can be expressed as
\begin{align*}
    \textbf{E}_{inc}^T(x, y, z) = \sum_{i, j}\textbf{E}_{inc}^T(x_i, y_j, z) f_{i, j}(x, y),
\end{align*}
where $x_i, y_j = i \lambda / 2, j \lambda / 2$ with $\lambda$ being the wavelength in the background medium, and $f_{i, j}(x, y)$ is a \emph{jinc} function \cite{goodman2005introduction} centered at $(x_i, y_j)$. Each term in the Nyquist decomposition can be considered to be an independent source, which falls off to zero with distance (Fig. \ref{fig:Fig2}a), and the response of a metasurface to these individual sources can be obtained by considering only a spatially-truncated portion of the metasurface in the simulation. This is numerically demonstrated in Fig. \ref{fig:Fig2}b, in which we consider the scattered power obtained on exciting a metasurface with a single jinc source as a function of the size of the metasurface included in the simulation. As the size of the metasurface is increased, the scattered power converges, indicating that a local simulation is sufficient to capture the metasurface response. The size of the metasurface to achieve a particular accuracy in the simulation is governed by diffraction of the jinc source by the time it reaches the metasurface.

\begin{figure*}[h!]
\centering
\includegraphics[width=\linewidth]{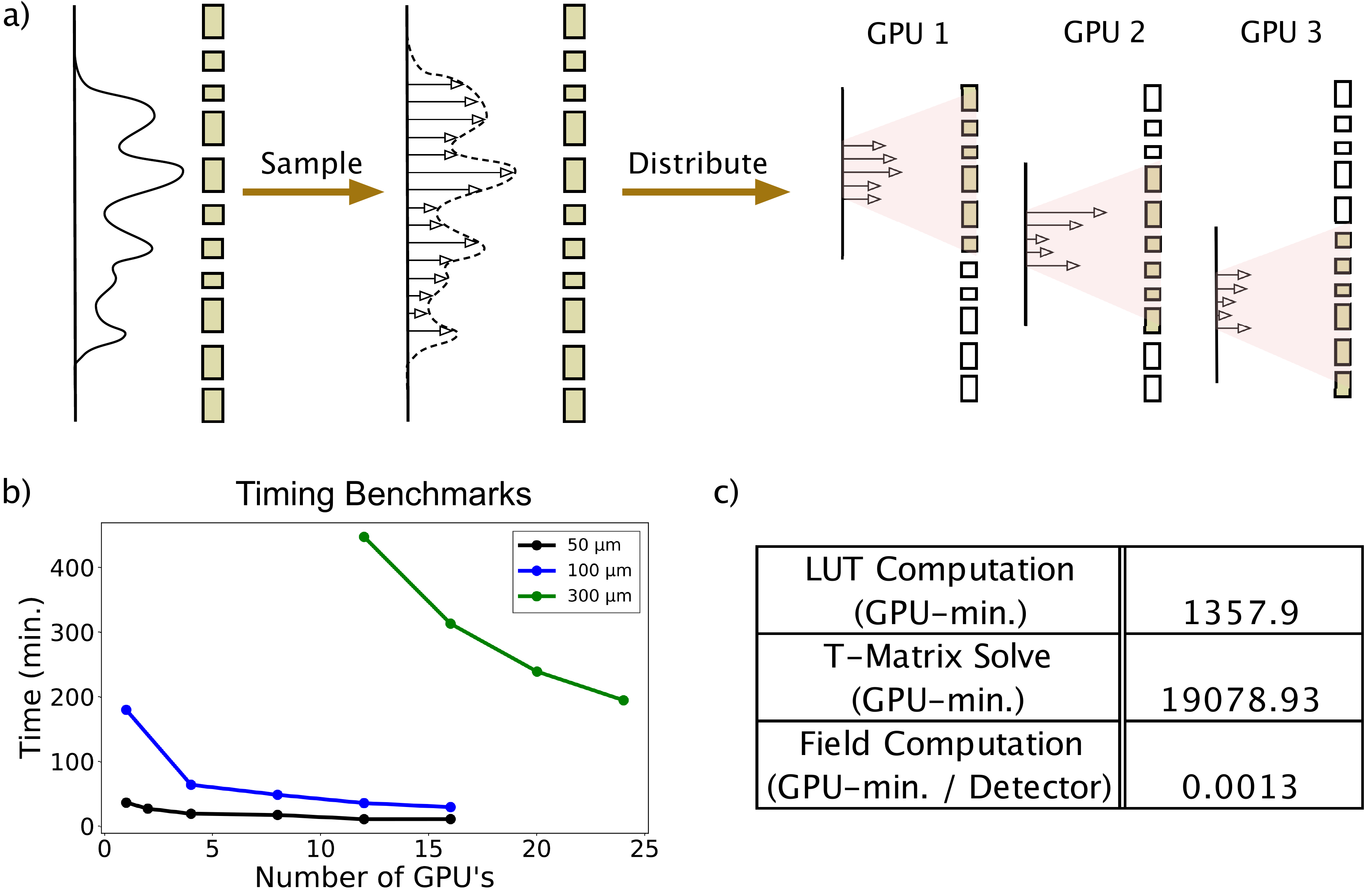}
\captionsetup{singlelinecheck=no}
\caption{{\bf{Low-overhead parallelization scheme to allow simulation of arbitrarily large metasurfaces.}} (\textbf{a}) Schematic of the simulation distribution scheme --- the incident field is first sampled and represented as a superposition of jinc sources, and then smaller groups of jinc sources and the locally surrounding metasurface regions are simulated on independent GPUs. (\textbf{b}) Total simulation time versus number of V100 GPU's used for simulation for a 50 $\mu$m (black), 100 $\mu$m (blue), and 300 $\mu$m (green) metasurface. All metasurfaces have focal length of $25 \mu$m and are designed from a library of silicon cylinders with height 940 nm, radii range of 50-250 nm, lattice period of 1070 nm, air background, and source wavelength of 1550 nm (based on scatterer library from\cite{arbabi2015subwavelength}). (\textbf{c}) Computation time for the key stages of the large-area 1 mm $\times$ 1 mm metasurface simulation: \textit{top row} -- computing the Look-Up Tables (LUT) used to efficiently perform T-matrix simulation (Appendix \ref{sec:tmat_details}); \textit{middle row} -- computing the T-matrices (Appendix \ref{sec:tmatrix}) and solving the resulting linear system of equations for the scattered field coefficients (Appendix \ref{sec:collective_scattering}, Eq. \ref{eq:final_linear_eqs}); \textit{bottom row} -- computing the E and H fields from the scattered field coefficients for each desired detector point (Appendix \ref{sec:collective_scattering}, Eq. \ref{eq:sca_fld_concise}). The simulated surface is a 1 mm $\times$ 1 mm metalens with focal length 0.4 mm (NA = 0.78) designed from a library of silicon cylinders with height 940 nm, radii range of 50-250 nm, lattice period of 1070 nm, air background, and source wavelength of 1550 nm (based on scatterer library from\cite{arbabi2015subwavelength}). The simulation is performed on 48 V100 GPUs and is distributed between these compute nodes using a subregion size of 20 $\mu$m $\times$ 20 $\mu$m and a padding of 6.5 $\mu$m, resulting in 2601 subregion simulations.}
\label{fig:FigDist}
\end{figure*}

To parallelize the simulation, we can then divide up the jinc sources that compose the incident electric fields into smaller groups, and simulate the local response of the metasurface for each source group by performing an independent solve on a single compute node (Fig.~\ref{fig:FigDist}a). This parallelization strategy only requires communication between the compute nodes at the start and end of the simulation --- once to distribute the simulation data corresponding to the local subregions, and then to consolidate the electric field data computed per subregion. On each compute node, we implement a GPU-parallelized transition-matrix (T-matrix) electromagnetic solver \cite{egel2017extending,doicu2006light} (See appendix \ref{sec:tmat_details} for details of the T-matrix method and our implementation of it). In order to accurately account for the diffraction of the jinc source while computing the local response of the metasurface, we include a padding region around the group of sources for each compute node. While in principle, we should ensure that the performance metric being analyzed (e.g.~metasurface efficiency) converges with respect to the padding, in practice and for typical metasurfaces, the thickness of padding required for accurate simulations can be estimated simply by studying the response of a local patch of the metasurface to one source similar to that done in Fig. \ref{fig:Fig2}b. After having performed all the simulations, the electric fields obtained can be added together to compute the total electric field. Furthermore, because each compute node performs roughly the same amount of compute, the total simulation time scales as ${1}/{N_{nodes}}$ (Fig. \ref{fig:FigDist}b). Details of our jinc source computation for the T-matrix method single-node simulation can be found in Appendix \ref{sec:jinc_source}.

Thus, given a sufficiently large number of compute nodes, we expect the simulation strategy to be able to handle large problems - on a compute cluster with 48 V100 GPU nodes, we were able to simulate metasurfaces of size about $645 \lambda \times 645 \lambda$ in about 10 hours. This total time is broken down into the compute times for the key simulation parts in Fig. \ref{fig:FigDist}c. \\

\begin{figure*}[h!]
\centering
\includegraphics[width=1.0\linewidth]{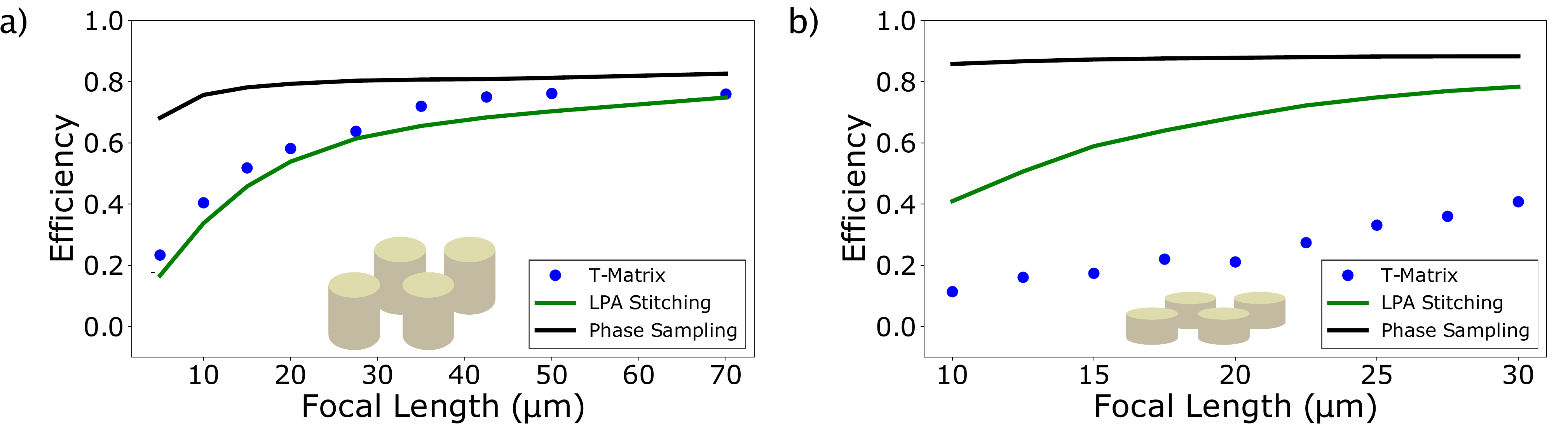}
\captionsetup{format=plain}
\caption{{\bf{Comparison of T-matrix method simulations with locally-periodic assumption (LPA) simulations.}} 
(\textbf{a}) Efficiency versus focal length for 25 $\mu$m $\times$ 25 $\mu$m metasurfaces designed from a library of high-aspect ratio scatterers with a large period (silicon cylinders with height 940 nm, radii range of 50-250 nm, lattice period of 1070 nm, and air background; source wavelength of 1550 nm -- based on scatterer library from \cite{arbabi2015subwavelength}) --- efficiencies are computed using the T-matrix approach (blue dots), the commonly-used LPA phase sampling approach (black curve), and the LPA field-stitching method (green curve). The metalens efficiency is defined as the ratio of the power within a circle of radius 3 $\times$ FWHM in the focal plane to the power incident on the metasurface. The T-matrix and LPA-stitching methods agree fairly well here because the scatterers are high-aspect ratio and the lattice constant is large, hence the interactions between neighboring scatterers is negligible. (\textbf{b}) Efficiency versus focal length for 15 $\mu$m $\times$ 15 $\mu$m metasurfaces designed from a library of low-aspect ratio scatterers with a small period (silicon cylinders with height 220 nm, radii range of 175-280 nm, lattice period of 666 nm, and background refractive index 1.66; source wavelength of 1340 nm -- using scatterer library from \cite{gigli2020fundamental}) --- efficiencies are computed using the T-matrix approach (blue dots), the commonly-used LPA phase sampling approach (black curve), and the LPA field-stitching method (green curve). The metalens efficiency is defined as the ratio of the power within a circle of radius 3 $\times$ FWHM in the focal plane to the power incident on the metasurface. The T-matrix and LPA-stitching methods do not agree here because the scatterers are low-aspect ratio and the lattice constant is small, hence the interaction between neighboring scatterers is significant.}
\label{fig:Fig3}
\end{figure*}

\noindent\textbf{Comparison with locally-periodic approximation.}
Approximate simulations of large-area metasurfaces often rely on the locally-periodic approximation (LPA) \cite{arbabi2015subwavelength,khorasaninejad2016polarization,pestourie2018inverse}, wherein the local electromagnetic response of the metasurface is approximated with that of a periodic metasurface. To demonstrate that the full metasurface simulation approach captures meta-atom interactions beyond LPA, we compare the T-matrix simulation method with two commonly-used LPA approaches in Fig. \ref{fig:Fig3}. First is a simple transmission mask approximation, wherein we assume that the metasurface imparts a smooth position-dependent amplitude and phase to the incident field as determined by the periodic simulation\cite{arbabi2015subwavelength}, and ignore the variation of the fields within a single unit cell. Second, we consider a more exact field stitching method \cite{li2021inverse}, wherein the fields near the metasurface within each unit cell is approximated with the fields from the periodic simulation and then propagated. For high aspect ratio scatterers, we find that while the transmission mask method significantly deviates from the T-matrix method, the field stitching method does not. However, for small aspect-ratio scatterers, which are expected to have larger inter meta-atom interactions, both the LPA approximations significantly deviate from the T-matrix method~\cite{gigli2020fundamental}. These results are a strong indication of the ability of the T-matrix method to capture meta-atom interactions and accurately simulate the metasurface response.

\begin{figure*}[t!]
\centering
\includegraphics[width=1.0\linewidth]{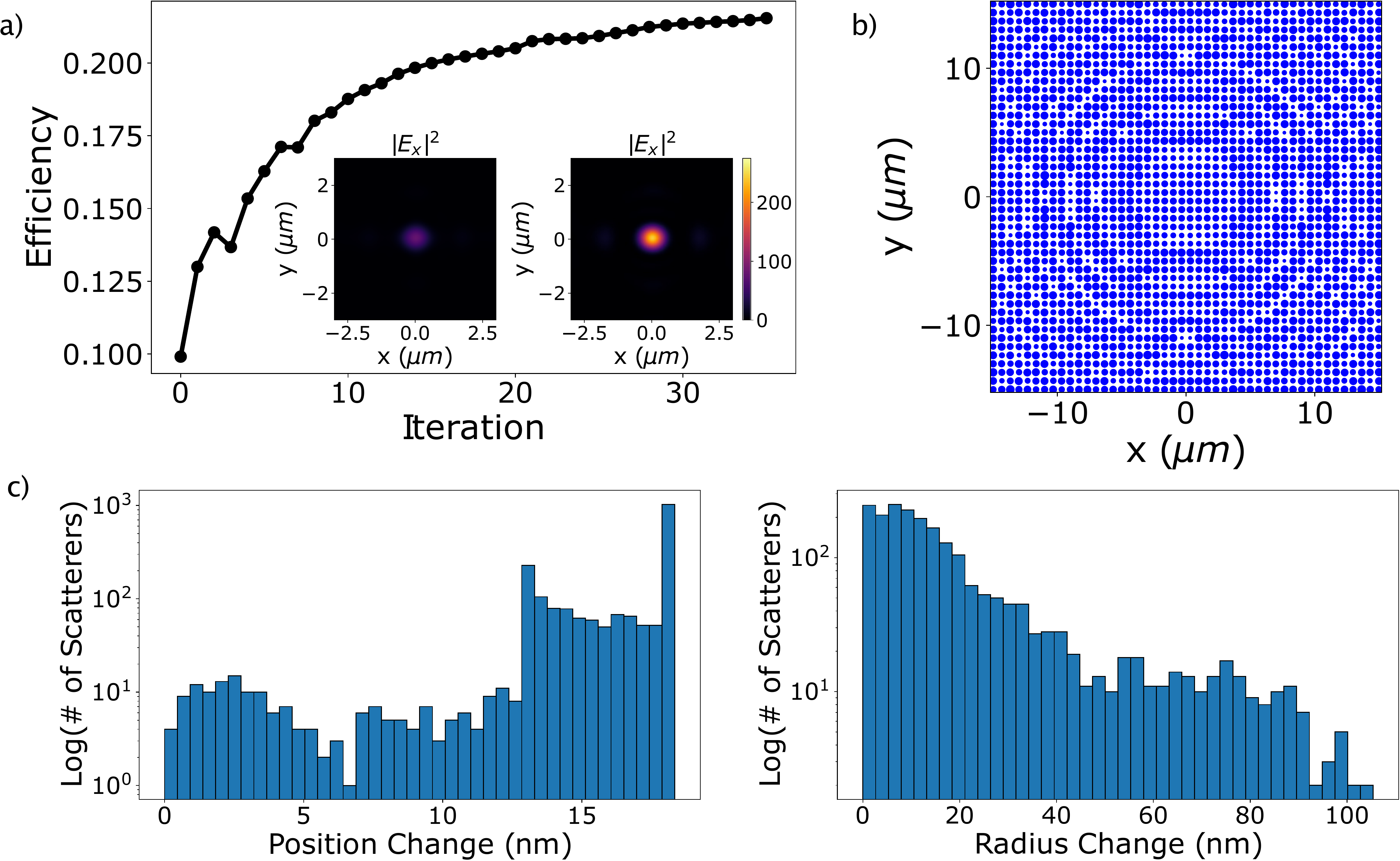}
\captionsetup{format=plain}
\caption{{\bf{Distributed Gradient-based optimization improvement of metalens design.}} 
(\textbf{a}) Lens efficiency versus optimization iteration, where lens efficiency is defined as the ratio of the power within a circle of radius 3 $\times$ FWHM in the focal plane to the power incident on the metasurface. The initial metasurface is a 30 $\mu$m $\times$ 30 $\mu$m metalens with focal-length 20 $\mu$m designed from the low-aspect ratio scatterer library in Fig. \ref{fig:Fig3}(b) using the traditional metasurface design approach. The metalens is  15 $\mu$m $\times$ 15 $\mu$m in size, and is optimized for x-polarized light only. In 35 optimization iterations, the metalens efficiency is almost doubled. The inset shows the X-component of the electric field in the focal plane before optimization (left) and after optimization (right). (\textbf{b}) Schematic of the cylindrical metasurface scatterers after optimization. (\textbf{c}) Histograms of the distance between the final scatterer positions and the initial scatterer positions (left) and the absolute radius difference between the final scatterer cylinders and the initial scatterer cylinders (right). As can be seen in these histograms, both the scatterer positions and radii change as a result of the optimization. }
\label{fig:Fig4}
\end{figure*}

\noindent\textbf{Distributed optimization-based design.}
An essential ingredient for optimization-based design of metasurfaces is an efficient evaluation of the gradient of the figure of merit with respect to the design parameters. A particularly useful method to evaluate gradients is based on adjoint-sensitivity analysis \cite{lalau2013adjoint,piggott2017fabrication} which analytically differentiates through Maxwell's equations and computes the gradients with respect to all the design parameters with a cost proportional to only two electromagnetic simulations. The distributed T-matrix simulation method is also amenable to distributed adjoint sensitivity analysis and can allow for scalable evaluation of the gradient of a performance metric defined on the electric fields scattered from the metasurface with respect to both the meta-atom shape and positions (see Appendix \ref{sec:adjoint_computation} for details). Fig. \ref{fig:Fig4} demonstrates a distributed gradient-based optimization with respect to the positions and radii of the cylindrical meta-atoms of a cost function evaluating the amount of power within a spot at the focal plane for a 30 $\mu$ m $\times$ 30 $\mu$m metalens with focal-length 20 $\mu$m initially designed with the same scatterer library used in Fig. \ref{fig:Fig3}(b). The distributed optimization was performed on 9 T4 GPUs with the metalens divided into 9 subregions (subregion size of 10 $\mu$m $\times$ 10 $\mu$ m, and padding size of 6 $\mu$ m). The forward simulations performed took an average of about 120 GPU-min and the gradient computations with respect to radius and position took an average of 150 GPU-min). The metalens has a very high NA of 0.996 and the optimization improves the efficiency of the metalens by about 2 $\times$, giving a final efficiency of about $24\%$. Although thin low-aspect ratio metasurfaces (Huygens metasurfaces) are of interest because they are more amenable to large-scale fabrication, they have not found widespread adoption due to their very limited efficiencies and angular responses \cite{gigli2020fundamental}. Our ability to accurately model the scatterer-scatterer effects in our metasurface inverse-design may allow discovery of more practical Huygens metasurfaces \cite{ollanik2018high,cai2020inverse}. Combining this multi-GPU gradient computation with the multi-GPU forward simulation, we have opened the door to gradient-based optimization over the many degrees of freedom afforded by arbitrarily large metasurfaces. In particular, our method allows optimizing both the shape and position of the scatterers composing the large-area metasurface --- optimizing the scatterer positions is very difficult for any inverse-design approach that relies on a periodicity assumption. \\

\noindent\textbf{Conclusion.}
We have demonstrated a scalable distribution method to accurately simulate arbitrarily large-area metasurfaces. Our method uses the Nyquist sampling theorem to allow parallel distribution of compute across multiple GPU nodes, on which a T-matrix method formulation is used to efficiently simulate the subregion. We show a roughly $\frac{1}{N_{GPU}}$ scaling of the total simulation time and demonstrate that our method accurately accounts for all scatterer interactions. Finally, we demonstrate our ability to apply our distribution method to the computation of the gradient with respect to all design parameters. Our simulation distribution method provides a solution to the long-standing problem of simulating large-area metasurfaces and opens the door to gradient-based optimization of the full metasurface, taking advantage of all the design degrees of freedom.\\

\clearpage
\noindent\textbf{Data availability} The data that support the plots within this paper and other findings of this study are available from the corresponding author upon reasonable request.\\

\noindent\textbf{Competing interests}
The authors declare that there are no competing interests.\\ \ \\

\noindent\textbf{Author Contributions}
R.T. and L.S. conceived the idea. J.S., R.T., and L.S. designed and implemented the software. J.S., R.T., L.S., D.A-S., and H.K. ran benchmarks and conducted numerical experiments. J.V., S.H., and S.F. supervised the project. All authors assisted with data analysis and manuscript preparation.\\ \ \\

\noindent\textbf{Acknowledgements} This work was supported by the Samsung GRO program. J.S acknowledges support from the National Science Foundation Graduate Research Fellowship (grant no. DGE-1656518) and Cisco Systems Stanford Graduate Fellowship (SGF). R.T acknowledges support from Max Planck Harvard research center for Quantum Optics (MPHQ) fellowship, and Sarah and Kailath Stanford Graduate Fellowship (SGF).

\bibliography{main_revised.bbl}
\clearpage

\onecolumngrid 

\appendix 
\renewcommand{\thefigure}{S\arabic{figure}}
\renewcommand{\thesection}{\Roman{section}}
\setcounter{figure}{0} 
\section* {Supplementary Information to\\Low-overhead distribution strategy for simulation and optimization of large-area metasurfaces}
\vspace{-0.15 in}
\noindent Jinhie Skarda$^{1,*}$, Rahul Trivedi$^{1,2,*,\dagger}$, Logan Su$^{1,*}$, Diego Ahmad-Stein$^{1}$, Hyounghan Kwon$^{1}$, \\Seunghoon Han$^{3}$, Shanhui Fan$^{1}$, and Jelena Vu\u{c}kovi\'{c}$^{1,\dagger}$
\vspace{0.1 in}\\
\noindent
$^1$E. L. Ginzton Laboratory, Stanford University, Stanford, CA 94305, USA.\\
$^2$Max-Planck-Institut für Quantenoptik, Hans-Kopfermann-Str.~1, 85748 Garching, Germany.\\
$^3$Samsung Advanced Institute of Technology, Samsung Electronics, Suwon-si, Gyeonggi-do 443-803, Republic of Korea\\
*These authors contributed equally to this work.\\
$^{\dagger}$Corresponding authors: rtriv@uw.edu,jela@stanford.edu\\

\section{T-Matrix Simulation Method}\label{sec:tmat_details}

\begin{figure*}[h!]
\centering
\includegraphics[width=\linewidth]{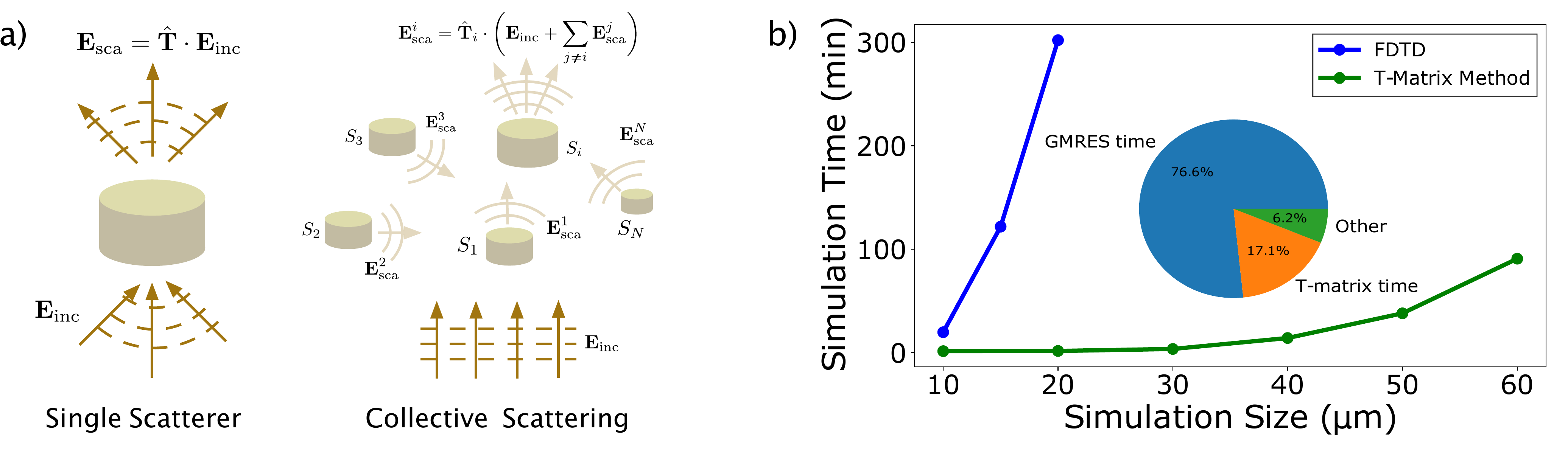}
\captionsetup{format=plain}
\caption{{\bf{Single-GPU T-matrix metasurface simulation method.}} (\textbf{a}) Schematic of T-matrix formulation for a single scatterer (left) and for collective scattering from multiple scatterers (right). (\textbf{b}) Simulation time versus simulation size for this single-GPU T-matrix method and for FDTD. The T-matrix method simulation was performed on a single V100 GPU, while the FDTD simulation was performed with Lumerical FDTD (Lumerical FDTD solutions, www.lumerical.com) on 8 CPUs with 32 GB RAM and mesh accuracy level 3. The inset breaks down the total simulation time for the 20 $\times$ 20 $\mu$m surface into the GMRES solve time (76.6\%), the time to compute the T-matrices (17.1\%), and all other computation (e.g. computing the incident field coefficients on the spherical harmonic basis functions and expanding the scattered field coefficients on the basis functions to compute the scattered electric field; 6.2\%).}
\label{fig:Fig1}
\end{figure*}

The key idea behind the T-matrix method is illustrated in Fig. \ref{fig:Fig1}(a)  --- the scattering properties of each meta-atom is captured by its T-matrix, which is a mathematical representation of the linear transformation mapping the incident field on the meta-atom to the scattered field. In principle, any basis for the incident or scattered fields can be used to represent this T-matrix. However, since the meta-atom sizes are of the order of a wavelength, choosing the spherical or spheroidal basis functions\cite{doicu2006light} provides a computationally economical description of the T-matrix (i.e.~only a modest number of basis functions are needed to capture the scattering properties of individual meta-atoms).

After computing the T-matrix for each meta-atom, a coupled system of equations can then be setup to account for interactions between different meta-atoms by treating the fields scattered from one meta-atom as being incident on the other meta-atoms. The coupling coefficients in such a system of equations are provided by the translation theorems for the chosen basis functions \cite{doicu2006light}, which allow an expansion for basis functions centered at one meta-atom on the basis functions centered at a different meta-atom. These coupled systems of equations are then solved to obtain a representation of the total scattered field on the chosen basis functions, which can then be used to compute the electric fields in position space and allow for an evaluation of any relevant metasurface performance metric.

To obtain a more computationally-efficient implementation of the T-matrix method, we implement several optimizations often used in the field of computational engineering. First, we use an iterative algorithm (GMRES) to solve the system of equations resulting from the T-matrix formalism --- a major advantage of using iterative algorithms is that they only require matrix-vector products which we program in a fully parallelized manner on GPUs. Furthermore, the computation of the T-matrix of the individual meta-atoms as well as of the coupling coefficients requires computing special functions which can be time consuming. In our implementation, we precompute lookup tables for these special functions before the start of the T-matrix simulation which significantly cuts down the run-time of the full simulation.

Fig.\ref{fig:Fig1}b shows a comparison of the simulation time of our implementation when compared to commercially available FDTD solvers \cite{solutions2003lumerical}. We note that this is not a strictly rigorous benchmark, as the parallelization settings are not exactly the same between our software and Lumerical -- the purpose of this comparison is only to roughly put our simulation times in context of a simulation tool the reader is likely familiar with. Our software is able to simulate a $60\times 60$ $\mu m^2$ large metasurface in 1-1.5 hours on a single GPU core (Nvidia V100). As seen in the inset in Fig. \ref{fig:Fig1}b, the bulk of the simulation time is spent in the linear system solve.

In the following subsections, we further describe the details of the T-matrix simulation method that we use on each compute node, and provide definitions of various objects used while setting up the simulations (e.g. vector spherical wave functions, transition matrices, translation theorems etc.). The focus is not derive the simulation method, but to present the equations that are ultimately implemented --- the interested reader can refer to \cite{Doicu} for a rigorous derivation of the simulation method.

\subsection{Vector spherical wavefunctions}
\noindent In this subsection, we provide definitions and some properties of the vector spherical wavefunctions. These wavefunctions serve as a basis functions to compactly represent the fields incident on and scattered from subwavelength scatterers, and are an important component of the transition matrix method. The definitions given here closely follow appendix B of \cite{Doicu}.

The vector spherical wavefunctions are solutions of the frequency-domain Maxwell's equations in homogenous media. Consider a medium with wavenumber $k = \omega \sqrt{\varepsilon} / c$ --- the electric field $\textbf{E}(\textbf{x})$ in source-free regions then satisfies:
\begin{align}\label{eq:vector_wf}
    \nabla \times \nabla \times \textbf{E}(\textbf{x}) - k^2 \textbf{E}(\textbf{x}) = 0
\end{align}
along with the transversality condition $\nabla \cdot \textbf{E}(\textbf{x}) = 0$ and the transversality condition $\nabla \cdot \textbf{E}(\textbf{x}) = 0$.

To construct the vector spherical wave functions, we begin by considering the solutions of the scalar helmholtz equation, which is the scalar counterpart of Eq.~\ref{eq:vector_wf}:
\begin{align}
    \nabla^2 \phi(\textbf{x}) + k^2 \phi(\textbf{x}) = 0
\end{align}
It can be shown, with a straightforward application of the separation of variables method, that the spherical wavefunctions $\phi_{l, m}(k_b\textbf{x})$ and $\mathcal{R}\phi_{l, m}(k_b\textbf{x})$ defined below are solutions of the scalar helmholtz equation:
\begin{subequations}
\begin{align}
    \phi_{l, m}(k\textbf{x}) = h_l^{(1)}(k r) P_l^{|m|}(\cos \theta) \exp(im \varphi) \\
    \mathcal{R}\phi_{l, m}(k\textbf{x}) = j_l(k r) P_l^{|m|}(\cos \theta) \exp(im \varphi)
\end{align}
\end{subequations}
where $l \in \{0, 1, 2 \dots \}$, $m \in \{-l, -l+1, \dots l-1, l\}$, $(r, \theta, \varphi)$ are the spherical coordinates of the point $\textbf{x}$, $h_l^{(1)}(x)$ is the spherical hankel function of the first kind, $j_l(x)$ is the spherical bessel function and $P_l^{|m|}(x)$ is the normalized associated legendre polynomial. Note that each solution of the scalar wave function is indexed by two numbers --- the orbital index $l$ and the magnetic index $m$. $\mathcal{R}$ indicates whether the wavefunction is well behaved at the origin --- $\mathcal{R}\phi_{l, m}(k\textbf{x})$ evaluates to 0 at the origin while $\phi_{l, m}(\textbf{x})$ diverges at the origin. Additionally, $\phi_{l,m}(k\textbf{x})$ captures spherical waves radiating to infinity, while $\mathcal{R}\phi_{l,m}(k\textbf{x})$ captures spherical standing waves (equal superpositions of outgoing and incoming waves) --- this can be best appreciated by considering the asymptotic forms of these wavefunctions at large radial coordinates ($k r \gg 1$):
\begin{align}\label{eq:asymptotic}
    &\phi_{l, m}(k\textbf{x}) \sim \frac{\exp(ik r)}{k r}P_{l}^{|m|}(\cos \theta) \exp(im\varphi) \\
    &\mathcal{R}\phi_{l, m}(k\textbf{x}) \sim \frac{[\exp(i(kr - l \pi /2) - \exp(-i(kr - l \pi / 2)) ]}{2ikr} P_l^{|m|}(\cos \theta)\exp(im\varphi)
\end{align}

For each scalar wave function, we can construct two vector spherical wave functions (corresponding to two independent polarizations) which satisfy Eq.~\ref{eq:vector_wf} along with the transversality condition $\nabla \cdot \textbf{E}(\textbf{x}) = 0$.  We denote the vector spherical wave functions by $\Phi_{l, m, p}(k_b \textbf{x})$ and $\mathcal{R}\Phi_{l, m, p}(k_b \textbf{x})$ (where $p \in \{0, 1\}$ is the polarization index) --- these are then defined by:
\begin{subequations}\label{eq:vector_spherical_wf}
\begin{align}
    &\begin{bmatrix} \Phi_{l, m, p=0}(k \textbf{x}) \\
                    \mathcal{R}\Phi_{l, m, p=0}(k \textbf{x})
    \end{bmatrix} = \frac{1}{\sqrt{2l (l+1)}} \nabla 
    \begin{bmatrix} \phi_{l, m}(k \textbf{x}) \\
                    \mathcal{R}\phi_{l, m}(k \textbf{x}) 
    \end{bmatrix} \times \textbf{x} \\
    &\begin{bmatrix} \Phi_{l, m, p=1}(k \textbf{x}) \\
                    \mathcal{R}\Phi_{l, m, p=1}(k \textbf{x})
    \end{bmatrix} = \frac{1}{k} \nabla \times 
    \begin{bmatrix} \Phi_{l, m, p=0}(k \textbf{x}) \\
                    \mathcal{R}\Phi_{l, m, p=0}(k \textbf{x}) 
    \end{bmatrix}
\end{align}
\end{subequations}
where $l \in  \{1, 2, 3 \dots\}$, $m \in \{-l, -l + 1 \dots l - 1, l \}$ and $p \in \{0, 1\}$. The vector spherical harmonics form a basis for the solutions of the vector helmholtz equation (Eq.~\ref{eq:vector_wf}), and an orthogonality condition can be constructed for them on the surface of (an arbitrarily chosen) sphere as delineated in \cite{Doicu}.

\subsection{Transition matrix}\label{sec:tmatrix}
\noindent Consider a scatterer with permittivity $\varepsilon_s$ embedded in a medium of permittivity $\varepsilon_b$ --- we denote by $\Gamma$ the volume of the scatterer, $\partial \Gamma$ the surface of the scatterer and by $\partial S$ the surface of a sphere enclosing the scatterer. Note that $\partial S$ can be chosen arbitrarily as long as it completely encloses the scatterer --- for performing practical computations, it is customary to choose $\partial S$ as the smallest sphere that can enclose the scatterer. Throughout this section, we assume that the center of $\partial S$ is at the origin of the coordinate system.

Consider exciting the scatterer with an incident electric field $\textbf{E}_{\text{inc}}(\textbf{x})$ that is a superposition of regular vector spherical wavefunctions $\mathcal{R}\Phi_{l, m, p}(k_b\textbf{x})$, where $k_b = \omega \sqrt{\varepsilon_b} / c$:
\begin{align}\label{eq:inc_fld_exp}
    \textbf{E}_{\text{inc}}(\textbf{x}) = \sum_{l, m, p} a_{l, m, p} \mathcal{R} \Phi_{l, m, p}(k_b \textbf{x})
\end{align}
This incident field interacts with the scatterer, resulting in a scattered field $\textbf{E}_{\text{sca}}(\textbf{x})$ radiating away from the scatterer. Outside the sphere $\partial S$, the scattered field can be expressed as a superposition of vector spherical wavefunctions $\Phi_{l, m, p}(k_b \textbf{x})$:
\begin{align}\label{eq:sca_fld_exp}
    \textbf{E}_{\text{sca}}(\textbf{x}) = \sum_{l, m, p}s_{l, m, p} \Phi_{l, m, p}(k_b \textbf{x})
\end{align}
Note that the scattered fields can be entirely captured without including the contribution of regular vector spherical wave functions --- this is a consequence of the fact that the scattered field is expected to propagate radially outward from the scatterer, and the regular vector spherical harmonics have a radially inward propagating component (Eq.~\ref{eq:asymptotic}). Moreover, the linearity of Maxwell's equations implies that the coefficients $s_{l, m, p}$ must be related to $a_{l, m, p}$ via a linear transformation:
\begin{align}\label{eq:def_tmatrix}
    s_{l, m, p} = \sum_{l', m', p'} T_{l, m, p; l', m', p'}a_{l', m', p'}
\end{align}
In practice, the expansions of the incident and scattered fields Eqs.~\ref{eq:inc_fld_exp} and \ref{eq:sca_fld_exp} are typically truncated to a finite number of terms by ignoring contributions of basis functions with $l > l_{\text{max}}$ [this corresponds to using $2l_{\text{max}}(l_{\text{max}}+2)$ basis functions]. In this case, all the coefficients $a_{l, m, p}$ ($s_{l, m, p}$) can be collected into a vector $\textbf{a}$ ($\textbf{s}$), and Eq.~\ref{eq:def_tmatrix} can be expressed as a matrix equation $\textbf{s} = \textbf{T}\textbf{a}$. We will refer to $\textbf{T}$ as the transition matrix of the scatterer.

There are a number of approaches to compute the transition matrix elements $T_{l, m, p; l', m', p'}$ --- we implement the approach based on the null field method \cite{Doicu}. The null field method uses the surface integral equations to individually relate the incident and scattered fields to the fields on the scatterer surface, expand all the surface fields on the vector spherical harmonic basis, and then compute the relationship between the incident and scattered fields by eliminating the surface fields from the resulting equations. Final computation of the transition matrix method involves computing two matrices $\textbf{P}$ and $\textbf{Q}$ whose elements are given by the following integrals:
\begin{subequations}
\begin{align}
&P_{l, m, p; l', m', p'} = \int_{\partial S}\bigg[\mathcal{R}\Phi_{l', m', p'}(k_s\textbf{x})\times \Phi_{l, -m, 1-p}(k_b \textbf{x}) - \frac{k_s}{k_b}\mathcal{R}\Phi_{l', m', 1-p'}(k_s\textbf{x})\times \Phi_{l, -m, p}(k_b \textbf{x})\bigg]\cdot \textbf{n}(\textbf{x}) \textrm{d}^2 \textbf{x} \\
&Q_{l, m, p; l', m', p'} = \int_{\partial S}\bigg[\mathcal{R}\Phi_{l', m', p'}(k_s\textbf{x})\times \mathcal{R}\Phi_{l, -m, 1-p}(k_b \textbf{x}) - \frac{k_s}{k_b}\mathcal{R}\Phi_{l', m', 1-p'}(k_s\textbf{x})\times \mathcal{R}\Phi_{l, -m, p}(k_b \textbf{x})\bigg]\cdot \textbf{n}(\textbf{x}) \textrm{d}^2 \textbf{x}
\end{align}
\end{subequations}
where $k_s$ is the wavenumber in the scatterer material, $k_b$ is the wavenumber in the background medium in which the scatterers are embedded, and the transition matrix $\textbf{T}$ is given by:
\begin{align}
    \textbf{T} = -\textbf{P} \textbf{Q}^{-1}
\end{align}
We use a numerical discretization of 0.0025 microns for evaluating these surface integrals to compute the T-matrix.

Finally, we note that the transition matrix becomes an increasingly accurate description of the scatterer on increasing $l_{\text{max}}$. However, for subwavelength scatterers, a small $l_{\text{max}}$ usually suffices to accurately describe the scattering properties of the scatterer. This is a key advantage of using the transition matrix method for simulating a collection of small scatterers, since the scattered fields from each scatterer can be described with a very small number of unknowns when expanded on the vector spherical wavefunctions. For all results in this paper, we have used $l_{\text{max}} = 6$.

\subsection{Simulating collective scattering}\label{sec:collective_scattering}
Finally, we describe how to simulate an ensemble of scatterers when their individual transition matrices are known. Consider $N$ scatterers with transition matrices $\textbf{T}_1, \textbf{T}_2 \dots \textbf{T}_N$ located at $\textbf{x}_1, \textbf{x}_2 \dots \textbf{x}_N$. We assume that the enclosing spheres ($\partial S$ defined in section \ref{sec:tmatrix}) of the scatterers do not intersect each other. This system of scatterers is excited with an incident field $\textbf{E}_{\text{inc}}(\textbf{x})$ which can be expanded into regular vector spherical wavefunctions centered at each of the scatterers:
\begin{align}
    \textbf{E}_{\text{inc}}(\textbf{x}) = \sum_{l, m, p} a^{(0)}_{l, m, p; i}\mathcal{R}\Phi_{l, m, n}(k_b(\textbf{x} - \textbf{x}_i))
\end{align}
where $i \in \{1, 2, 3 \dots N\}$ and $k_b$ is the wavenumber of the background medium. Note that the coefficients $a_{l, m, n; i}^{(0)}$ for different $i$ are not independent of each other --- in particular, given these coefficients for one choice of $i$, the coefficients for all other $i$ can be constructed via an application of the translation theorem for vector spherical wavefunctions \cite{Doicu}. Typically, the incident field is known analytically (e.g.~plane wave field), and hence these coefficients can be analytically determined for an arbitrary $i$.

We express the scattered field as a superposition of vector spherical harmonics radiated from each scatterer:
\begin{align}\label{eq:sca_fld_sys}
    \textbf{E}_{\text{sca}}(\textbf{x}) = \sum_{i=1}^N \sum_{l, m, p}s_{l, m, p; i} \Phi_{l, m, p}(k_b(\textbf{x} - \textbf{x}_i))
\end{align}
where it is assumed that $\textbf{x}$ lies outside the enclosing spheres of all scatterers. Consider now the $j^{\text{th}}$ scatterer --- the total incident field at this scatterer, $\textbf{E}_{{\text{inc}}, j}(\textbf{x})$ is given by the sum of $\textbf{E}_{\text{inc}}(\textbf{x})$ as well as fields radiated by the neighbouring scatterers:
\begin{align}
    \textbf{E}_{{\text{inc}},j}(\textbf{x}) = \textbf{E}_{\text{inc}}(\textbf{x}) + \sum_{\substack{i=1 \\ i \neq j}}^N \sum_{l, m, p} s_{l, m, p; i} \Phi_{l, m, p}(k_b (\textbf{x} - \textbf{x}_i))
\end{align}
This field can be expressed entirely as a superposition of the regular vector spherical wavefunctions centered at $\textbf{x}_j$ by the application of the translation theorem for the vector spherical wavefunctions. The translation theorem relates vector spherical wavefunctions defined with respect to two different origins to each other \cite{Doicu,Egel} --- more explicitly:
\begin{align}
    \Phi_{l, m, p}(k(\textbf{x} - \textbf{x}_a)) = \sum_{l', m', p'} \xi_{l, m, p; l', m', p'}(k(\textbf{x}_a - \textbf{x}_b))\Phi_{l', m', p'}(k(\textbf{x} - \textbf{x}_b))
\end{align}
where
\begin{align}
    \xi_{l, m, p; l', m', p'}(\textbf{x}) = \delta_{p, p'} \alpha_{l, m; l', m'}(\textbf{x}) + (1 - \delta_{p, p'}) \beta_{l, m; l', m'}(\textbf{x})
\end{align}
and with $(r, \varphi, \theta)$ as the spherical coordinates of the vector $\textbf{x}$:
\begin{subequations}
\begin{align}
    &\alpha_{l, m; l', m'}(\textbf{x}) = \exp(i(m - m')\varphi)\sum_{q = |l-l'|}^{l + l'} a_5(l, m | l', m' | q)h_q^{(1)}(r) P_q^{|m - m'|}(\cos \theta) \\
    &\beta_{l, m; l', m'}(\textbf{x}) = \exp(i(m - m')\varphi)\sum_{q = |l-l'| + 1}^{l + l'} b_5(l, m | l', m' | q)h_q^{(1)}(r) P_q^{|m - m'|}(\cos \theta)
\end{align}
\end{subequations}
where:
\begin{subequations}\label{eq:tran_coeff}
\begin{align}
\alpha(l, m|l', m'|p) = &i^{|m - m'| - |m|-|m'| + l' - l + p}(-1)^{m - m'} \\ \nonumber
                        &\times [l(l + 1) + l'(l' + 1) - p(p+1)]\sqrt{2p+1} \\ \nonumber
                        &\times \sqrt{\frac{(2l + 1)(2l' + 1)}{2l(l+1)(l'+1)}}
                        \begin{pmatrix}
                        l & l' & p \\
                        m & -m' & m' - m
                        \end{pmatrix}
                        \begin{pmatrix}
                        l & l' & p \\
                        0 & 0 & 0
                        \end{pmatrix} \\
\beta(l, m|l', m'|p) = &i^{|m - m'| - |m|-|m'| + l' - l + p}(-1)^{m - m'} \\ \nonumber
                        &\times \sqrt{(l + l' + 1 + p)(l + l' + 1 - p)(p + l - l')(p - l + l')(2p + 1)} \\ \nonumber
                        &\times \sqrt{\frac{(2l + 1)(2l' + 1)}{2l(l+1)(l'+1)}}
                        \begin{pmatrix}
                        l & l' & p \\
                        m & -m' & m' - m
                        \end{pmatrix}
                        \begin{pmatrix}
                        l & l' & p \\
                        0 & 0 & 0
                        \end{pmatrix}
\end{align}
\end{subequations}
with
\begin{align}
    \begin{pmatrix} j_1 & j_2 & j_3\\
                    m_1 & m_2 & m_3
    \end{pmatrix}
\end{align}
is the Wigner-3j symbol \cite{Luscombe}. With this addition theorem, $\textbf{E}_{\text{inc},j}(\textbf{x})$ can be rewritten as:
\begin{align}
    \textbf{E}_{\text{inc}, j}(\textbf{x}) = \sum_{l, m, p} a_{l, m, p; j} \mathcal{R}\Phi_{l, m, p}(k_b (\textbf{x} - \textbf{x}_j))
\end{align}
where
\begin{align}
    a_{l, m, p; j} = a^{(0)}_{l, m, p; j} + \sum_{\substack{i = 1 \\ i \neq j}}^N\sum_{l', m', p'}\xi_{l', m', p'; l, m, p}(k_b(\textbf{x}_j - \textbf{x}_i))s_{l', m', p'; i}
\end{align}
Finally, the coefficients $a_{l, m, p; j}$ can be related to the coefficients $s_{l, m, p; j}$ via the transition matrix of the $j^{\text{th}}$ scatterer: $\textbf{s}_j = \textbf{T}_j\textbf{a}_j$, where $\textbf{s}_j$ and $\textbf{a}_j$ are column vectors of the coefficients $s_{l, m, n; j}$ and $a_{l, m, n; j}$. Denoting by $\boldsymbol{\xi}(\textbf{x}_i, \textbf{x}_j)$ as the matrix of the translation coefficients $\xi_{l, m, p; l', m', p'}(k_b(\textbf{x}_i - \textbf{x}_j))$ with the unprimed indices corresponding to different rows and the primed indices corresponding to different columns, this equation can be compactly written as:
\begin{align}
    \textbf{T}_j^{-1} \textbf{s}_j - \sum_{\substack{i=1 \\ i \neq j}}^N \boldsymbol{\xi}^T(\textbf{x}_j, \textbf{x}_i) \textbf{s}_i = \textbf{a}^{(0)}_j
\end{align}
Such an equation can be written for each scatterer --- all of these equations collected together provide a completely determined system of linear equations that can be solved to obtain the scattering coefficients $s_{l, m, p; j}$. The system of equations can be written in a matrix form:
\begin{align}\label{eq:final_linear_eqs}
    \underbrace{\begin{bmatrix}
    \textbf{T}_1^{-1} & \boldsymbol{\xi}^T(\textbf{x}_1, \textbf{x}_2) & \boldsymbol{\xi}^T(\textbf{x}_1, \textbf{x}_3) & \dots & \boldsymbol{\xi}^T(\textbf{x}_1, \textbf{x}_N) \\
    \boldsymbol{\xi}^T(\textbf{x}_2, \textbf{x}_1) & \textbf{T}_2^{-1} & \boldsymbol{\xi}^T(\textbf{x}_2, \textbf{x}_3) & \dots & \boldsymbol{\xi}^T(\textbf{x}_2, \textbf{x}_N) \\
    \boldsymbol{\xi}^T(\textbf{x}_3, \textbf{x}_1) & \boldsymbol{\xi}^T(\textbf{x}_3, \textbf{x}_2) & \textbf{T}_3^{-1} & \dots & \boldsymbol{\xi}^T(\textbf{x}_3, \textbf{x}_N) \\
    \vdots                   & \vdots                   & \vdots            & \ddots & \vdots                  \\
    \boldsymbol{\xi}^T(\textbf{x}_N, \textbf{x}_1) & \boldsymbol{\xi}^T(\textbf{x}_N, \textbf{x}_2) & \boldsymbol{\xi}^T(\textbf{x}_N, \textbf{x}_3) & \dots & \textbf{T}_N^{-1}
    \end{bmatrix}}_{\boldsymbol{\Omega}}
    \underbrace{\begin{bmatrix}
    \textbf{s}_1 \\
    \textbf{s}_2 \\
    \textbf{s}_3 \\
    \vdots \\
    \textbf{s}_N
    \end{bmatrix}}_{\textbf{s}} =
    \underbrace{
    \begin{bmatrix}
    \textbf{a}^{(0)}_1 \\
    \textbf{a}^{(0)}_2 \\
    \textbf{a}^{(0)}_3 \\
    \vdots \\
    \textbf{a}^{(0)}_N
    \end{bmatrix}}_{\textbf{a}^{(0)}}
\end{align}
The matrix $\boldsymbol{\Omega}$ in the above system of equations is referred to as the `maxwell operator`, since it is equivalent to expressing the frequency domain maxwell's equations on the vector spherical wavefunction basis. To summarize, the problem of solving the Maxwell's equations has been reduced to the problem of solving the linear system in Eq.~\ref{eq:final_linear_eqs}. Once this system of equations is solved to obtain $s_{l, m, n;j}$, the electric fields at the desired points can be computed using Eq.~\ref{eq:sca_fld_sys}, which can be concisely expressed as:
\begin{align}\label{eq:sca_fld_concise}
    \textbf{E}_{\text{sca}}(\textbf{x}) = \sum_{i=1}^N \textbf{s}_i^T \Phi(\textbf{x} - \textbf{x}_i)
\end{align}
where $\Phi(\textbf{x})$ is a column vector of the vector spherical harmonics $\Phi_{l, m, p}(\textbf{x})$ at position $\textbf{x}$.

\begin{figure}
    \centering
    \includegraphics[width=\linewidth]{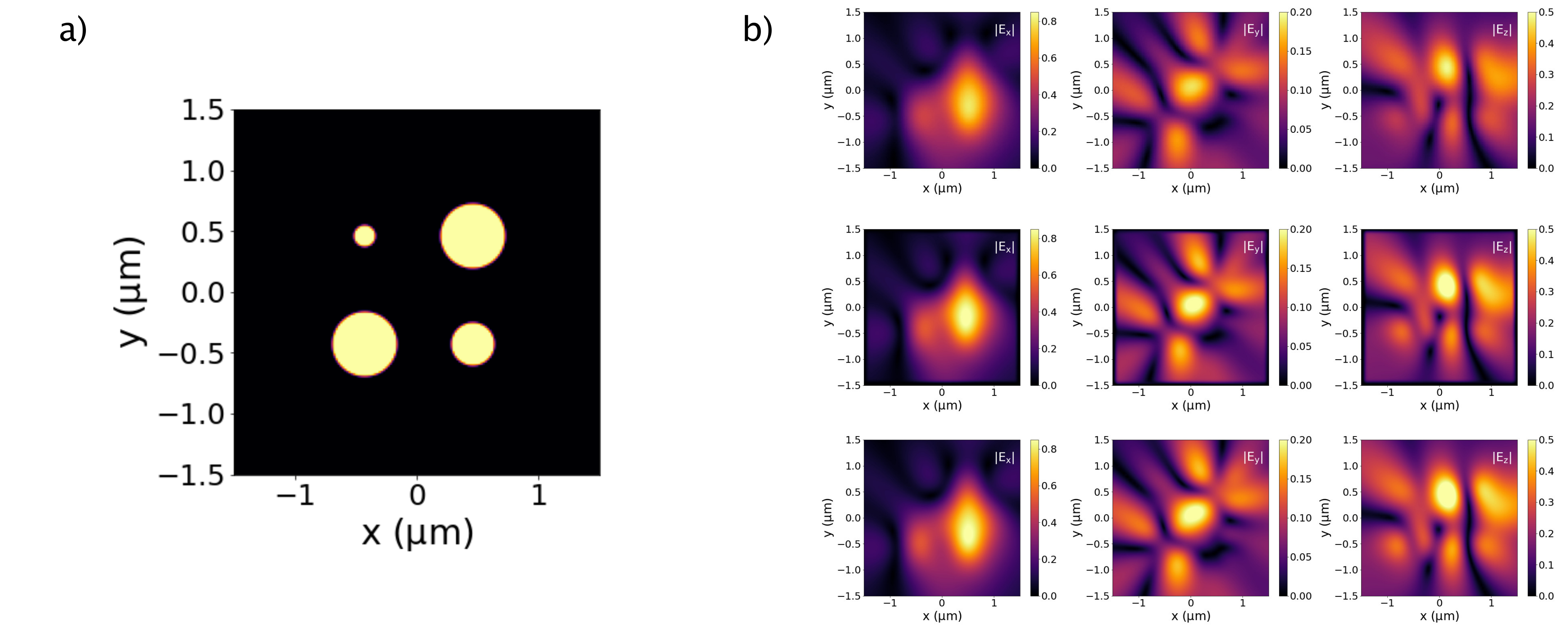}
    \caption{(a) Scatterers used for the validation simulation --- the scatterers are illuminated with a plane wave and have a refractive index of 3.5. (b) Electric field magnitudes for the x, y, and z components of the scattered fields from the T-matrix method simulation (top), FDFD simulation (middle), and FDTD simulation (bottom).}
    \label{fig:validation}
\end{figure}

From a numerical perspective, the linear systems Eq.~\ref{eq:final_linear_eqs} is severely ill conditioned due to the transition matrices being close to singular for small scatterers. This issue can be mitigated using the following block diagonal preconditioner $\textbf{M}$:
\begin{align}
    \textbf{M} = \begin{bmatrix}
    \textbf{T}_1 &  &  &  \\
    & \textbf{T}_2 &  & \\
    & & \textbf{T}_3 & &  \\
    & & & \ddots & \\
    & & & & \textbf{T}_N  \\
    \end{bmatrix}
\end{align}
Instead of solving $\boldsymbol{\Omega} \textbf{s} = \textbf{a}^{(0)}$, we solve $\textbf{M}\boldsymbol{\Omega}\textbf{s} = \textbf{M}\textbf{a}^{(0)}$ (a system of equations similar to those found in \cite{markkanen2017fast}). Validation simulations for our T-matrix simulation method implementation are shown in Fig. \ref{fig:validation} --- we see good agreement between the T-matrix method, FDTD, and FDFD simulations.

\subsection{Hardware acceleration on GPUs}\label{sec:single_gpu}
\begin{figure}
    \centering
    \includegraphics[scale=0.3]{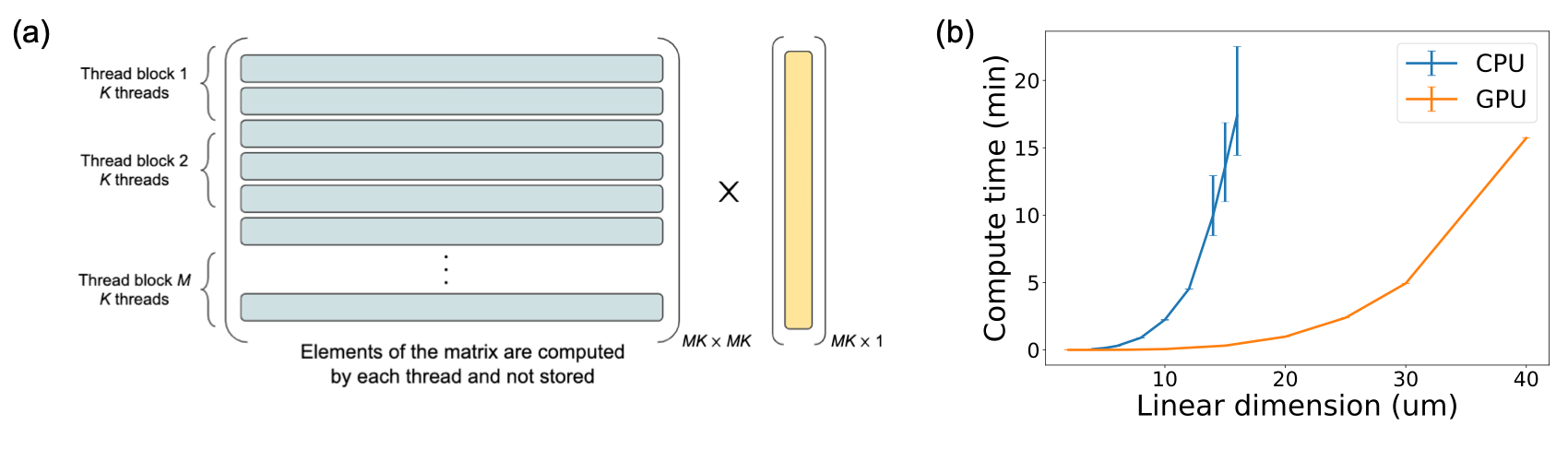}
    \caption{(a) Schematic of the distribution of a matrix-vector product across a GPU --- each thread (grouped as thread blocks) is assigned to perform the computation of the product between one row and the vector with the thread computing any matrix-elements required and discarding them once the computation is done. (b) Comparison of the solve time between the CPU and GPU implementations for a 2D array of cylinders located at randomly chosen positions within a rectangle of the specified linear dimensions. The error bars indicate the spread in the solve time in between 10 different randomly chosen configurations of the cylinders for the same linear dimension. Note that both GPU and CPU simulations are performed with GMRES with a residual of $10^{-6}$. All the GPU simulations were performed on GTX Titan Black with 6GB memory.}
    \label{fig:gpu_cpu}
\end{figure}

\noindent Accelerating the solution of the resulting system of linear equations using hardware accelerators such as GPUs has been an immensely successful approach in scaling up partial differential equation solvers. In this section, we describe a simple approach to implement the transition matrix simulation on GPUs --- in particular, there are two issues that we address using common approaches from the field of computational engineering:
\begin{enumerate}
    \item \emph{Speeding up matrix solve}: The matrix solve using GMRES relies on the ability to perform matrix vector products (i.e.~if we are solving $\textbf{A}\textbf{x} = \textbf{b}$ using GMRES, we need to be able to compute the product of the matrix $\textbf{A}$ with an arbitrary vector $\textbf{x}$). For a $N \times N$ dense matrix, this computation is an $O(N^2)$ operation, and consequently speeding up this operation is a key component of scaling up the simulator to larger systems.
    \item \emph{Memory obstacle in precomputing the full matrix}: The implementation described in the previous section constructs the full maxwell operator explicitly as a matrix, and then performs the matrix vector products. While this might work for small scale simulations ($\approx 15 \ \mu$m in linear dimension for subwavelength silicon scatterers while using a machine with $8$ GM RAM), for larger simulations the matrix would become too large to store in memory. So as to obviate this issue, we would like to be able to perform matrix-vector products without explicitly constructing the matrix.
\end{enumerate}
\textbf{Implementation of the GPU accelerated solver}: To this end, we implement the matrix vector product with the maxwell operator as a GPU operation (Fig.~\ref{fig:gpu_cpu}a). The matrix elements are computed while performing the matrix-vector product and discarded once the computation involving those elements have been performed. While computing the $k^{\text{th}}$ element of the matrix-vector product, the inner product of the $k^{\text{th}}$ row of the matrix with the vector needs to be computed --- in our implementation, we assign one GPU thread to handle one such inner product so as to parallelize the matrix-vector product operation. This matrix-vector product operation can then be used along with GMRES to fully solve the system of equations --- we implement the operations in GMRES other than the matrix-vector product operation using nvidia's CUBLAS library \cite{cublas}. Fig.~\ref{fig:gpu_cpu}b shows a comparison between the matrix solve time between a CPU implementation of the solution of Eq.~\ref{eq:final_linear_eqs} and a GPU implementation of the solution of Eq.~\ref{eq:final_linear_eqs}. Both the implementations use lookup tables and interpolation --- note that in the CPU solve time, we include the time taken for constructing the maxwell operator and then solving Eq.~\ref{eq:final_linear_eqs} using GMRES. We observe a $10\times$ speedup on using the GPU implementation over the CPU implementation. \\

\section{Computation of Jinc Sources}\label{sec:jinc_source}
While the single-GPU implementation of the T-matrix method outlined in Section \ref{sec:single_gpu} is efficient for metasurfaces with linear dimension up to $30 \mu m$, it is not scalable to millimeter scale metasurface simulations as can be seen by an extrapolation of the timing benchmark shown in Fig.~\ref{fig:gpu_cpu}b. In this section, we describe an approach to perform the metasurface simulation over multiple GPUs with very little communication overhead, thereby allowing us to scale the T-matrix method to large-scale metasurface simulations.

The fundamental principle behind the parallelization scheme that we implement is the Nyquist sampling theorem. Consider an incident field propagating along the $+z$ axis and described by its transverse components as a function of transverse coordinate $(x, y)$ at $z = 0$, $\textbf{E}^{T}_{\text{inc}}(x, y, z = 0)$ is incident on a metasurface located at $z > 0$. If the incident field is produced by a source that is either far from the metasurface or paraxial, it will be spatially bandlimited to a light-cone. Consequently, at $z = 0$, it can be completely described by its samples $\textbf{E}^t(x_i, y_j, z=0)$ where $x_i = i \lambda_0 / 2$ and $y_j = j\lambda_0 / 2$:
\begin{align}
\textbf{E}^{T}_{\text{inc}}(x, y, z = 0) = \sum_{i,j=-\infty}^\infty  \frac{j_1(k_0 \rho_{i,j})}{k_0 \rho_{i,j}} \textbf{E}^{T}_{\text{inc}}(x_i, y_j, z = 0)
\end{align}
where $j_1(\cdot)$ is the spherical bessel function of order 1 and $\rho_{i,j} = \sqrt{(x - x_i)^2 + (y - y_j)^2}$. Thus with every sample, one can associate an incident field, labelled as `jinc source', with amplitude proportional to the sampled electric field. It can easily be seen that the jinc source is of limited spatial extent, and consequently simulating the response of the metasurface to an individual jinc source only requires performing a simulation of a small part of the metasurface near its center. Consequently, we can divide up the jinc sources that compose the incident electric fields into small groups, and simulate the response of the metasurface locally for each group by performing an independent solve on single GPU. After having performed all the simulations, they can be added together again to compute the total electric field.

We note that, although we have made the assumption that the incident field is propagating at normal incidence, this Nyquist sampling technique also works for oblique incidence. Keeping the same sampling plane as for normal incidence, the Nyquist sampling rate will increase because the maximum magnitude of the in-plane k-vector increases. As long as the angle of incidence is close to normal, this sampling strategy should have performance comparable to the normal-incidence case.

\subsection{Implementing jinc source in the vector spherical harmonic basis}
In order to implement the Nyquist-sampling-based-parellelization scheme described above, it is necessary to be able to simulate the response of the metasurface to a collection of jinc sources using the T-matrix method. This requires the ability to expand the jinc sources on the vector spherical wavefunctions (Eq.~\ref{eq:final_linear_eqs}). In this section, we mathematically develop such an expansion.

Consider a jinc source at $z = z_0$ propagating in the $+z$ direction and centered at $\vx^t_0 = (x_0, y_0)$ in the transverse plane. The transverse electric field at $z = z_0$ is given by:
\begin{align}
\textbf{E}^T(\textbf{x}^T, z = z_0) = \textbf{A}^T \frac{j_1(k_0|\textbf{x}^T - \textbf{x}^T_0|)}{k_0 |\textbf{x}^T - \textbf{x}^T_0|}
\end{align}
where $j_1(\cdot)$ is the first order spherical bessel function and $\textbf{A}^T = A_x \hat{x} + A_y \hat{y}$ is the transverse polarization of the jinc field. Alternatively, in the fourier representation,
\begin{align}
    \textbf{E}^T(x, y, z = z_0) = \frac{\textbf{A}^T}{2 \pi} \int_{|\textbf{k}^T| < k_0}\exp[i \textbf{k}^T\cdot (\textbf{x}^T - \textbf{x}^T_0)]d^2\textbf{k}^T
\end{align}
Propagating each plane-wave component, we obtain:
\begin{align}
\textbf{E}(x, y, z) = \frac{1}{2\pi}\int_{|\textbf{k}^T| < k_0} \textbf{A}(\textbf{k}^T) \exp(i\textbf{k}^T \cdot (\textbf{x}^T - \textbf{x}^T_0)) \exp(ik_z (z - z_0)) {d}^2\textbf{k}^T 
\end{align}
where $k_z = \sqrt{k_0^2 - \textbf{k}^T\cdot\textbf{k}^T}$ and
\begin{align}
\textbf{A}(\vk^T) = \textbf{A}^T - \hat{z}\frac{\vk^T \cdot \textbf{A}^T}{k_z(\vk^T)}
\end{align}

Changing the integration variable to $\alpha, \beta$ ($0\leq \alpha \leq 2 \pi$ and $0 \leq \beta \leq \pi / 2$) where $\textbf{k}^T(\beta, \alpha) = k_0\sin \beta (\hat{x} \cos \alpha + \hat{y} \sin \alpha)$ and therefore $k_z(\vk^T)\equiv k_z(\beta, \alpha) = k_0 \cos \beta$, $\textbf{A}(\textbf{k}^T) \equiv \textbf{A}(\beta, \alpha)= \hat{x} A_x + \hat{y} A_y - \hat{z} \tan \beta (A_x \cos \alpha + A_y \sin \alpha)$ and $d^2 \vk^T = k_0^2 \sin\beta \cos \beta d\alpha d\beta$. Moreover, each plane wave can be expanded into a sum of vector spherical harmonic wavefunctions \cite{Doicu}:
\begin{align}
    &\textbf{A}(\beta, \alpha) \exp(i\vk^T(\beta, \alpha) \cdot (\vx^T - \vx^T_0)) \exp(ik_z(\beta, \alpha) (z - z_0)) = \exp[i\vk(\beta, \alpha) \cdot (\vx_0' - \vx_0)]\sum_{l, m, p} a_{l, m, p}(\beta, \alpha) \mathcal{R}\vPhi_{l, m, p}(k_0 (\vx - \vx_0'))
\end{align}
where
\begin{subequations}
\begin{align}
    &a_{l, m, p=0}(\beta, \alpha) = 4i^m \textbf{A}(\beta, \alpha)\cdot \textbf{m}_{l, m}^*(\beta, \alpha)=-\frac{4i^l \exp(-im\alpha)}{\sqrt{2l(l+1)}}\big[im \pi^{|m|}_l(\beta) (\hat{\beta}\cdot\textbf{A}(\beta, \alpha)) + \tau^{|m|}_l(\beta) (\hat{\alpha}\cdot \textbf{A}(\beta, \alpha)) \big] \label{eq:integrand_p_0}\\
    &a_{l, m, p=1}(\beta, \alpha) = -4i^{l+1} \textbf{A}(\beta, \alpha) \cdot \textbf{n}_{l, m}^*(\beta, \alpha) = -\frac{4i^{l+1}\exp(-im\alpha)}{\sqrt{2l(l+1)}}\big[\tau^{|m|}_l(\beta) (\hat{\beta}\cdot\textbf{A}(\beta, \alpha)) - im\pi^{|m|}_l(\beta) (\hat{\alpha}\cdot \textbf{A}(\beta, \alpha)) \big]\label{eq:integrand_p_1}
\end{align}
\end{subequations}
where $\textbf{m}_{l, m}(\cdot, \cdot)$ and $\textbf{n}_{l, m}(\cdot, \cdot)$ are vector spherical harmonics with orbital index $l$ and azimuthal index $m$. Therefore,
\begin{align}
    \vE(\vx) = \sum_{l, m, p} a_{l, m, p} \mathcal{R}\vPhi_{l, m, p}(k_0 (\vx - \vx_0'))
\end{align}
where $a_{l, m, p}$ are given by:
\begin{align}
    a_{l, m, p} = \frac{1}{2\pi}\int_{\beta=0}^{\pi / 2}\int_{\alpha=0}^{2\pi} a_{l, m, p}(\beta, \alpha) \exp[i\vk(\beta, \alpha)\cdot (\vx_0' - \vx_0)]\sin \beta \cos \beta d\alpha d\beta
\end{align}
In the remainder of this section, we will evaluate the integral over $\alpha$ analytically, and the resulting expression can then be numerically integrated over $\beta$. To do so, it will be convenient to define a function $\Gamma_m(\xi, \eta, \rho)$ by:
\begin{align}
    \Gamma_m(\xi, \eta, \rho) = \frac{1}{2\pi}\int_0^{2\pi} \exp(-im \alpha) \exp[i\rho \cos(\alpha - \xi)]\cos(\alpha - \eta) d\alpha
\end{align}
$\Gamma_{m}(\xi, \eta, \rho)$ can be evaluated analytically: making a change of variables to $\alpha' = \alpha - \xi + \pi / 2$, we obtain:
\begin{align}
\Gamma_m(\xi, \eta, \rho) &= \frac{i^m\exp(-im \xi)}{2\pi} \int_{\pi/2 - \xi}^{5\pi / 2 - \xi} \exp(-im \alpha') \exp(i\rho\sin \alpha') \sin(\alpha' + \xi -\eta)d\alpha' \nonumber \\
    &=\frac{i^{m-1}\exp(-im \xi)}{4\pi}\int_0^{2\pi}\exp(-im\alpha')\exp(i\rho\sin \alpha')(\exp(i(\alpha' + \xi - \eta)) - \exp(-i(\alpha' + \xi - \eta)))d\alpha'\nonumber \\
    &=\frac{i^{m-1}\exp(-im\xi)}{2}\bigg[\exp\{i(\xi - \eta)\} J_{m-1}(\rho) - \exp\{-i(\xi - \eta)\} J_{m+1}(\rho) \bigg]
\end{align}
wherein we have used the identity:
\begin{align}
    J_m(\rho) = \frac{1}{2\pi}\int_0^{2\pi} \exp[i\rho \sin \theta - m \theta]d\theta
\end{align}
Additionally, note that $\hat{\beta} =  (\hat{x}\cos \alpha + \hat{y}\sin \alpha)\cos \beta - \hat{z}\sin \beta $ and $\hat{\alpha} = -\hat{x}\sin \alpha + \hat{y}\cos \alpha$. Therefore:
\begin{subequations}
\begin{align}
    &\hat{\beta}\cdot \textbf{A}(\beta, \alpha) = \sec \beta (A_x \cos \alpha + A_y \sin \alpha) \\
    &\hat{\alpha} \cdot \textbf{A}(\beta, \alpha) = -A_x \sin \alpha + A_y \cos \alpha
\end{align}
\end{subequations}
Finally, let $\vx_0' - \vx_0 \equiv (r_0, \theta_0, \varphi_0)$, and therefore $\vk(\beta, \alpha) \cdot (\vx_0' - \vx_0) = k_0 r_0 (\cos \beta \cos \theta_0 + \sin \beta \sin \theta_0 \cos(\alpha - \varphi_0 ))$.

\begin{enumerate}
    \item Consider the computation of $a_{l, m, p = 0}$. Using Eq.~\ref{eq:integrand_p_0}, we obtain:
    \begin{align}
        a_{l,m, p=0}(\beta, \alpha)\sin \beta \cos \beta= -\frac{4i^l \exp(-im\alpha)}{\sqrt{2l(l+1)}}\big[&(im\pi_l^{|m|}(\beta) A_x + \tau_l^{|m|}(\beta) \cos \beta A_y)\cos \alpha \nonumber \\ &+ (im \pi^{|m|}_l(\beta) A_y - \tau_l^{|m|}(\beta) \cos \beta A_x)\sin \alpha \big] \sin \beta
    \end{align}
    and therefore
    \begin{align}
        \frac{1}{2\pi}\int_0^{2\pi}a_{l, m, p=0}(\beta, \alpha) \exp[i\vk(\beta, \alpha) \cdot (\vx_0 -\vx_0')]\sin \beta \cos \beta d\alpha = -\frac{4i^l\exp(ik_0 r_0 \cos \beta \cos \theta_0)\sin \beta}{\sqrt{2l(l+1)}}\times\nonumber \\
        \bigg[(im\pi_l^{|m|}(\beta) A_x + \tau_l^{|m|}(\beta) \cos \beta A_y)\Gamma_m(\phi_0, 0, k_0 r_0\sin \beta \sin \theta_0) +\nonumber \\  (im \pi^{|m|}_l(\beta) A_y - \tau_l^{|m|}(\beta) \cos \beta A_x)\Gamma_m(\phi_0, \pi / 2, k_0 r_0 \sin \beta \sin \theta_0)\bigg]
    \end{align}
    
    \item Consider the computation of $a_{l, m, p = 1}$. Using Eq.~\ref{eq:integrand_p_1}, we obtain:
    \begin{align}
        a_{l, m, p=1}(\beta, \alpha) \sin \beta \cos \beta =-\frac{4i^{l+1}\exp(-im\alpha)}{\sqrt{2l (l + 1)}}\big[(\tau_l^{|m|}(\beta) A_x - im \pi_l^{|m|}(\beta) A_y \cos \beta)\cos \alpha \nonumber \\
                 +(\tau_l^{|m|}(\beta) A_y + im \pi_l^{|m|}(\beta) A_x \cos \beta)\sin \alpha\big]
    \end{align}
    and therefore
    \begin{align}
        \frac{1}{2\pi}\int_0^{2\pi} a_{l, m, p=1}(\beta, \alpha)\exp[i\vk(\beta, \alpha) \cdot (\vx_0 - \vx_0')] \sin \beta \cos \beta d\alpha = -\frac{4i^{l+1} \exp(ik_0 r_0 \cos \beta \cos \theta_0) \sin \beta}{\sqrt{2l(l+1)}}\times\nonumber \\\big[(\tau_l^{|m|}(\beta) A_x - im \pi_l^{|m|}(\beta) A_y \cos \beta)\Gamma_m(\phi_0, 0, k_0r_0 \sin \beta \sin \theta_0) \nonumber \\+(\tau_l^{|m|}(\beta) A_y + im \pi_l^{|m|}(\beta) A_x \cos \beta)\Gamma_m(\phi_0, \pi/2, k_0 r_0 \sin \beta \sin \theta_0)\big]
    \end{align}
\end{enumerate}

It can be noted that the numerical integration over $\beta$ can be accelerated by using lookup tables for the various special functions involved in the computation. Furthermore, we also parallelize the computation of the jinc sources on GPU to accelerate it with one thread being assigned to compute $a_{l, m, p}$ for a single choice of $(l, m, p)$ with respect to a chosen scatterer.

Finally, we remark that since the jinc sources are spatially limited, we only need to simulate the metasurface locally to compute its response to the jinc source. In order to estimate how local this simulation needs to be, a padding study like the one in Fig. \ref{fig:Fig2}(b) should be performed.

\section{Scatterer Libraries for Metalens Designs}\label{sec:scat_libraries}
The transmission and phase response for the scatterer library used for the metalens designs simulated and analyzed in Fig. \ref{fig:Fig2}, \ref{fig:Fig3}(a), and \ref{fig:Fig4} is shown below in Fig.\ref{fig:scat_library} (a). The transmission and phase response for the scatterer library used for the metalens designs simulated and analyzed in Fig. \ref{fig:Fig3}(b) is shown below in Fig.\ref{fig:scat_library} (b).

\begin{figure}
    \centering
    \includegraphics[width=\linewidth]{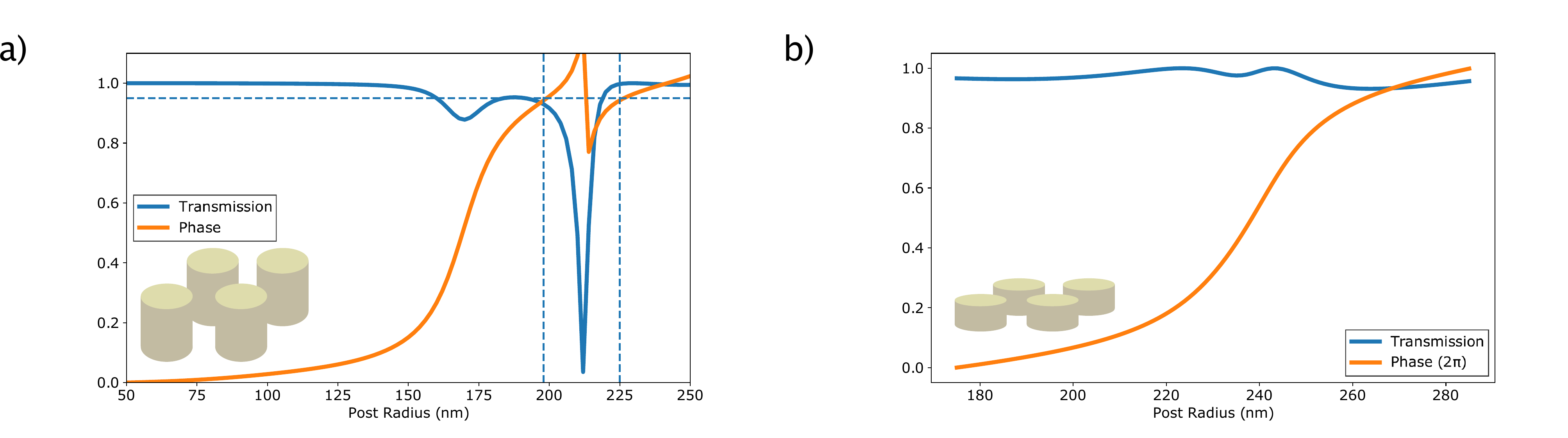}
    \caption{(a) Transmission and phase response for the scatterer library based on\cite{Arbabi}, consisting of silicon cylinders with height 940 nm, radii range of 50-250 nm, square lattice period of 1070 nm, air background, and plane wave source wavelength of 1550 nm. (b) Transmission and phase response for the scatterer library based on\cite{Gigli}, consisting of silicon cylinders with height 220 nm, radii range of 175-280nm, square lattice period of 666 nm, background refractive index of 1.66, and plane wave source wavelength of 1340 nm.}
    \label{fig:scat_library}
\end{figure}

\section{Adjoint Computation}\label{sec:adjoint_computation}
\noindent In this section, we describe the computation of the gradient of the metasurface performance with respect to the position and geometry of the meta-atoms. An efficient implementation of the gradient computation would allow us to use gradient-based optimization algorithms to optimize the performance of metasurfaces much like that done with inverse-design of silicon photonics devices.

Suppose that we have $N$ scatterers located at $\textbf{x}_1, \textbf{x}_2 \dots \textbf{x}_N$. Moreover, the geometry of the scatterers are dependent on parameters $p_1, p_2 \dots p_M$ (e.g.~these parameters can be the radii of the cylindrical meta-atoms, or the lengths and breadths rectangular meta-atoms). Consider a performance metric $\mathcal{O}$ which takes the following form:
\begin{align}\label{eq:performance_metric}
    \mathcal{O} = f\bigg(\int_{\Gamma_d}\boldsymbol{\sigma}^*(\textbf{x})\cdot \textbf{E}_{\text{sca}}(\textbf{x})\textrm{d}^3\textbf{x} \bigg)
\end{align}
where $f(\cdot)$ is a function mapping a complex number to a real number, $\Gamma_d$ is the detector volume and $\boldsymbol{\sigma}(\textbf{x})$ is a vector function of space that has information of the `desired' field profile. As a concrete example of such a performance metric, consider attempting to design a lens using meta-atoms located at $z = 0$ that focuses an incident plane-wave polarized along $\hat{\textbf{v}}$ within a Gaussian spot of width $w_0$ at $z = F$. In this case, we could choose $f(z) = |z|^2$ and  $\boldsymbol{\sigma}(\textbf{x}) = \hat{\textbf{v}}\exp[-(x^2 + y^2)/w_0^2] \delta(z - F)$ so as to obtain a performance metric that measures how much of the transmitted field is within a gaussian beam mode of the specified width $w_0$ at the focus of the metasurface. More complicated performance metrics can be created by combining those of the form of Eq.~\ref{eq:performance_metric} based on the specific design problem being solved.

Consider now the computation of derivative of $\mathcal{O}$ with respect to the coordinates of the $i^{\text{th}}$ scatterer: $\partial \mathcal{O} / \partial q_i$ where $q_i \in \{x_i, y_i, z_i\}$ with $\textbf{x}_i = (x_i, y_i, z_i)$. From Eq.~\ref{eq:performance_metric}, it immediately follows that:
\begin{align}
    \frac{\partial \mathcal{O}}{\partial q_i} = 2 \textrm{Re}\bigg[ f'\bigg(\int_{\Gamma_d} \boldsymbol{\sigma}^*(\textbf{x})\cdot \textbf{E}_{\text{sca}}(\textbf{x}) \textrm{d}^3\textbf{x} \bigg) \frac{\partial}{\partial q_i} \bigg(\int_{\Gamma_d} \boldsymbol{\sigma}^*(\textbf{x})\cdot \textbf{E}_{\text{sca}}(\textbf{x})\textrm{d}^3\textbf{x}\bigg)\bigg]
\end{align}
From Eq.~\ref{eq:sca_fld_concise}, it immediately follows that:
\begin{align}
    \frac{\partial}{\partial q_i}\bigg(\int_{\Gamma_d} \boldsymbol{\sigma}^*(\textbf{x})\cdot \textbf{E}_{\text{sca}}(\textbf{x}) \textrm{d}^3\textbf{x} \bigg) = \textbf{v}^{\textrm{T}}\frac{\partial}{\partial q_i}\begin{bmatrix}
    \textbf{s}_1 \\
    \textbf{s}_2 \\
    \vdots \\
    \textbf{s}_N
    \end{bmatrix}+ \textbf{s}_i^{\textrm{T}} \bigg(\int_{\Gamma_d}\boldsymbol{\sigma}^*(\textbf{x})\cdot\frac{\partial}{\partial q_i}\Phi(\textbf{x} - \textbf{x}_i)\textrm{d}^3\textbf{x}\bigg)
\end{align}
where
\begin{align}
    \textbf{v} = \int_{\Gamma_d}\boldsymbol{\sigma}^*(\textbf{x})\cdot\begin{bmatrix}
    \Phi(\textbf{x} - \textbf{x}_1) \\
    \Phi(\textbf{x} - \textbf{x}_2)\\
    \vdots \\
    \Phi(\textbf{x} - \textbf{x}_N)
    \end{bmatrix}\textrm{d}^3\textbf{x}
\end{align}
To compute $\partial \textbf{s}_k /\partial q_i$, we differentiate Eq.~\ref{eq:final_linear_eqs} with respect to $q_i$:
\begin{align}
    \frac{\partial}{\partial q_i}\begin{bmatrix}
    \textbf{s}_1 \\
    \textbf{s}_2 \\
    \vdots \\
    \textbf{s}_N
    \end{bmatrix}= -\boldsymbol{\Omega}^{-1} \frac{\partial \boldsymbol{\Omega}}{\partial q_i} 
    \begin{bmatrix}
    \textbf{s}_1 \\
    \textbf{s}_2 \\
    \vdots \\
    \textbf{s}_N
    \end{bmatrix}
\end{align}

from which we obtain:
\begin{align}\label{eq:grad_pos_overlap}
    \frac{\partial}{\partial q_i}\bigg(\int_{\Gamma_d}\boldsymbol{\sigma}^*(\textbf{x})\cdot\textbf{E}_{\text{sca}}(\textbf{x})\textrm{d}^3\textbf{x}\bigg) = -\textbf{v}^{\textrm{T}} \boldsymbol{\Omega}^{-1} \frac{\partial \boldsymbol{\Omega}}{\partial q_i} 
    \begin{bmatrix}
    \textbf{s}_1 \\
    \textbf{s}_2 \\
    \vdots \\
    \textbf{s}_N
    \end{bmatrix} + \textbf{s}_i^{\textrm{T}} \bigg(\int_{\Gamma_d}\boldsymbol{\sigma}^*(\textbf{x})\cdot\frac{\partial}{\partial q_i}\Phi(\textbf{x} - \textbf{x}_i)\textrm{d}^3\textbf{x}\bigg)
\end{align}

Following a similar procedure, it is possible to compute the partial derivative of $\mathcal{O}$ with respect to the geometric parameter $p_i$ to obtain:
\begin{align}
    \frac{\partial \mathcal{O}}{\partial p_i} = 2\text{Re}\bigg[f'\bigg(\int_{\Gamma_d}\boldsymbol{\sigma}^*(\textbf{x})\cdot\textbf{E}_{\text{sca}}(\textbf{x})\textrm{d}^3\textbf{x}\bigg)\frac{\partial }{\partial p_i}\bigg(\int_{\Gamma_d}\boldsymbol{\sigma}^*(\textbf{x})\cdot\textbf{E}_{\text{sca}}(\textbf{x})\textrm{d}^3\textbf{x}\bigg) \bigg]
\end{align}
where
\begin{align}\label{eq:grad_geometry_overlap}
    \frac{\partial}{\partial p_i} \bigg(\int_{\Gamma_d} \boldsymbol{\sigma}^*(\textbf{x})\cdot \textbf{E}_{\text{sca}}(\textbf{x}) \textrm{d}^3\textbf{x}\bigg) = -\textbf{v}^{\textrm{T}} \boldsymbol{\Omega}^{-1}\frac{\partial \boldsymbol{\Omega}}{\partial p_i}    \begin{bmatrix}
    \textbf{s}_1 \\
    \textbf{s}_2 \\
    \vdots \\
    \textbf{s}_N
    \end{bmatrix}
\end{align}

From Eqs.~\ref{eq:grad_pos_overlap} and \ref{eq:grad_geometry_overlap}, we make the following observations about the gradient computation:
\begin{enumerate}
    \item[(a)] It seems that computing the gradient requires computation of $\boldsymbol{\Omega}^{-1}$ --- this would be a prohibitively expensive simulation that would render the gradient computation impractical. Note however, in the gradient computation, we only require the computation of $\textbf{v}^{\text{T}}\boldsymbol{\Omega}^{-1} = [\boldsymbol{\Omega}^{-\text{T}}\textbf{v}]^{\text{T}}$ and not the full inverse $\boldsymbol{\Omega}^{-1}$. This is equivalent to solving the following system of equations:
    \begin{align}
        \boldsymbol{\Omega}^{\text{T}}\textbf{a} = \textbf{v}
    \end{align}
    This is labelled as the adjoint simulation, and it needs to be performed \emph{once} at each step of the optimization (importantly, note that this simulation does \emph{not} depend on the variable with respect to which the gradient is being computed).
    
    \item[(b)] To compute the gradients, we also need to compute $\partial \boldsymbol{\Omega} / \partial q_i$ and $\partial \boldsymbol{\Omega} / \partial p_i$. Since the explicit dependence of $\boldsymbol{\Omega}$ on the scatterer coordinates is known (i.e.~ Eq.~\ref{eq:final_linear_eqs}), $\partial \boldsymbol{\Omega} / \partial q_i$ can be computed analytically. Computing $\partial \boldsymbol{\Omega}/\partial p_i$ is equivalent to computing $\partial \textbf{T}^{-1}_k / \partial p_i$ (note that the off-diagonal elements do not depend on the scatterer geometry, only on their locations). Since in our implementation, we are setting up a lookup table for the transition matrices and interpolating between then, this derivative can be numerically approximated using the same lookup table.
\end{enumerate}

\end{document}